\documentclass[9pt,aps,pra,twocolumn]{revtex4-1}
\usepackage{amsmath,amssymb,graphicx,comment}
\usepackage{color}
\usepackage{bm}
\usepackage{hyperref}
\usepackage{bbm}
\usepackage{tabularx,ragged2e,booktabs}

\newcommand{\eq}[1]{\begin{align}#1\end{align}}

\newcommand{\seq}[1]{\begin{subequations}#1\end{subequations}}
\newcommand{\sseq}[1]{\seq{\eq{#1}}}

\newcommand{\V}{\bm{V}}
\DeclareMathOperator{\one}{\mathbbm{1}}

\definecolor{JM}{RGB}{4,116,149}

\definecolor{TB}{RGB}{144,177,52}

\definecolor{BG}{RGB}{144,1,52}

\definecolor{blue}{rgb}{0,0.2,1}

\definecolor{red}{rgb}{0.9,0,0}

\begin{document}
\title{Classical benchmarking of Gaussian Boson Sampling on the Titan supercomputer}

\author{Brajesh Gupt}
\email{brajesh@xanadu.ai}
\author{Juan Miguel Arrazola}
\author{Nicol\'as Quesada}
\author{Thomas R. Bromley}
\affiliation{Xanadu, 372 Richmond St W, Toronto, M5V 2L7, Canada}

\begin{abstract}
Gaussian Boson Sampling is a model of photonic quantum computing where
single-mode squeezed states are sent through linear-optical interferometers and
measured using single-photon detectors.  In this work, we employ a recent exact
sampling algorithm for GBS with threshold detectors to perform classical
simulations on the Titan supercomputer. We determine the time and memory
resources as well as the amount of computational nodes required to produce
samples for different numbers of modes and detector clicks.  It is possible to
simulate a system with 800 optical modes postselected on outputs with 20
detector clicks, producing a single sample in roughly two hours using $40\%$ of the available nodes of Titan. Additionally,
we benchmark the performance of GBS when applied to dense subgraph
identification, even in the presence of photon loss. We perform sampling for
several graphs containing as many as 200 vertices. Our findings indicate that
large losses can be tolerated and that the use of threshold detectors is
preferable over using photon-number-resolving detectors postselected on
collision-free outputs.  
\end{abstract}

\maketitle

\section{Introduction}
The first generation of programmable quantum devices is emerging. This has led
to an increased interest to understand their practical applications and their
potential to surpass the capabilities of traditional computers
\cite{preskill2018quantum,harrow2017quantum}. Classical simulation algorithms
play an important role in this development: they can be used to benchmark the
correctness of quantum algorithms and to set the bar of performance for quantum
computers \cite{pednault2017breaking, chen201864, zulehner2018advanced,
biamonte2018quantum, chen2018classical}. 

In photonic quantum computing, boson sampling is a sub-universal model where
indistinguishable single photons are sent through linear optics interferometers
and their output ports are recorded using single-photon detectors
\cite{aaronson2011computational,spring2012boson,broome2013photonic,tillmann2013experimental}.
Despite its conceptual simplicity, it is believed that simulating the behavior
of a boson sampling device requires exponential time on a classical computer
\cite{aaronson2011computational, aaronson2013bosonsampling}. This standard boson
sampling model requires single-photon sources, which are challenging to realize
experimentally at a large scale. Consequently, other variants of boson sampling
have been proposed where the inputs are squeezed states, which are
more amenable to implement in practice. Examples of these models include scattershot boson
sampling  \cite{lund2014boson,bentivegna2015experimental,latmiral2016towards}
and Gaussian Boson Sampling (GBS) \cite{hamilton2017gaussian,kruse2018detailed}. Notably, it has been shown that GBS has applications in quantum
chemistry \cite{huh2015boson,clements2017experimental,sparrow2018simulating},
optimization \cite{arrazola2018using, arrazola2018quantum}, and graph theory
\cite{bradler2017gaussian}.

Alongside these theoretical and experimental developments, significant progress
has been made in designing classical algorithms for simulating boson sampling.
In Ref.~\cite{neville2017classical}, a Markov chain Monte Carlo algorithm for
approximate boson sampling was introduced, capable of significantly
outperforming strategies based on a brute force calculation of the probability
distribution. This result was improved in Ref.~\cite{clifford2018classical},
where an exact boson sampling algorithm was proposed having the same asymptotic
complexity as the algorithm of Ref.~\cite{neville2017classical}, but lower
runtime for fixed problem sizes. 

Both of these algorithms rely on special
properties of the matrix permanent, which characterizes the probability
distribution of standard boson sampling with single-photon inputs. These algorithms  have not been extended to other boson sampling models, whose
probability distributions are described by different matrix functions. Instead,
Ref.~\cite{quesada2018gaussian} introduced a physically-motivated and exact
classical algorithm for GBS with threshold detectors. These are detectors that
register a click when one or more photons are observed, but are incapable of
resolving photon number. The GBS distribution with threshold detectors
approximates conventional GBS with photon-number-resolving detectors when the
probability of more than one photon reaching a given port is small, i.e., so
that samples are collision free.

In this work, we employ the classical algorithm of
Ref.~\cite{quesada2018gaussian} to perform classical simulation of GBS
with threshold detectors. We determine the time and memory resources as
well as the amount of computational nodes required to produce samples for
different numbers of optical modes and detector clicks. The computational
resources required to implement the algorithm increase exponentially with the
number of detector clicks and polynomially with the number of modes. Therefore,
for large system sizes, the simulation becomes intractable for traditional
desktop computers. For such simulations, we employ the Titan supercomputer
housed at Oak Ridge National Laboratories. We employ distributed memory
parallelization using the message passing interface (MPI) protocol together with
OpenMP multi-threading for increased performance, thus fully exploiting the
capabilities of the supercomputer. This allows us to push the limits of a full
simulation of GBS for a system of 800 modes, postselecting on outcomes with 20
detector clicks using 240,000 CPU cores, producing a single sample in roughly two
hours. 

Additionally, we employ our simulations to study the performance of a recently
proposed method of using GBS for dense subgraph identification
\citep{arrazola2018using}. We first study a small graph of 30 vertices with a
planted densest subgraph of 10 vertices. Our simulation results show that using
threshold detectors significantly increases the performance compared to
photon-number-resolving detectors that are postselected on collision-free
samples. This indicates that threshold detectors, besides being experimentally
appealing due to their low cost and room-temperature operation, can lead to
better results for specific applications. We also simulate sampling from this graph in the presence of photon loss. We find that significant improvement compared to classical sampling remains even in the presence of losses as large as 6~dB. Finally, we use our simulations to
find dense subgraphs of a graph of 200 nodes from the DIMACS dataset
\cite{johnson1996cliques}. 

The paper is organized as follows. In section \ref{sec:algo}, we briefly review the
algorithm introduced in Ref.~\cite{quesada2018gaussian} for GBS with threshold
detectors. In section \ref{sec:bench}, we analyze the memory and runtime
requirements of the algorithm and discuss the benchmarking results using the
Titan supercomputer to produce samples of various system sizes. In section
\ref{sec:app}, we employ the simulations for dense subgraph identification. We conclude with a discussion of our results in section \ref{sec:disc}.

\section{The Algorithm} \label{sec:algo}
In this section, we detail the classical algorithm introduced in Ref.~\cite{quesada2018gaussian} to simulate GBS with threshold detectors. Consider a GBS device with $\ell$ modes. The main idea of the algorithm is that, even though the
positive-operator valued measure (POVM) element representing a click is
non-Gaussian, it can be written as the difference of two Gaussian
operators. Thus, whenever the POVM element corresponding to a
click is applied to a Gaussian state, it is possible to represent the
conditional state of the remaining modes as the difference of two Gaussian states. Given the linearity
of the action of the measurements on a quantum state, the aforementioned result generalizes to a linear combination of several Gaussian states when more clicks are detected. 

Before presenting the algorithm in detail, we introduce useful notation
related to Gaussian states. An $\ell$-mode Gaussian state is uniquely
characterized by a $2\ell \times 2 \ell$ covariance matrix $\bm{V}$ and a vector
of means $\bm{\bar r}$. These quantities are defined as the following
expectation values on the Gaussian state $\rho$:
\sseq{
\bm{V}_{ij} &= \frac{1}{2} \langle \Delta \hat r_i \Delta \hat r_j+ \Delta \hat r_i \Delta \hat r_j \rangle_{\rho},\\
\Delta \hat{\bm{ r}}&=\hat{ \bm{r}}-\bar{ \bm{r}},\\
\bar{ \bm{r}} &= \langle \hat{\bm{r}} \rangle_{\rho} = \text{Tr}\left(\hat{ \bm{r}}  \rho \right),\\
\bm{\hat r} &= (\hat x_1,\hat p_1,\ldots,\hat x_\ell,\hat p_\ell)^T~,
}
where $\hat x_j$ and $\hat p_k$ are the canonical quadratures of the modes
satisfying the commutation relation $[\hat x_j, \hat p_k] = 2 i \delta_{j,k}$, where we have set $\hbar=2$.
Having set up the notation for covariance matrices, we write $\rho(\bm{V},
\bm{\bar r})$ to identify the Gaussian state with covariance matrix $\bm{V}$ and
vector of means $\bm{\bar r}$. 

The algorithm is based on a sequence of measurements on the output ports of
the device. Each time a mode is measured, the new state is computed via
the following update rule. Upon measuring the $\ell^{\text{th}}$ mode of an $\ell$-mode 
state that is a linear combination of $M$ Gaussian states, 
we obtain an $\ell-1$ mode state. If a click is recorded, the new state can be
expressed as a linear combination of $2M$ Gaussian states. If no click is
recorded, it is a linear combination of $M$ Gaussian states. Thus, each click gives rise to a doubling of the number of states that must be recorded. Sequential
measurements on all modes then gives a string of click events, namely a
measurement sample. More precisely, the update rule proceeds as follows: 
\begin{itemize}
\item Input: an $\ell$-mode state 
\eq{			\rho_{\ell} = {\sum_{k=1}^{M}} a_k \rho_{\ell,k}(\bm{V}_k,\bm{\bar r}_k),}
that is a linear combination of Gaussian states $ \rho_{\ell,k}(\bm{V}_k,\bm{\bar r}_k)$.
\item For each element of the linear combination $k$, decompose the covariance matrix as  
\eq{
	\V_k &{\to} \begin{bmatrix}
		\noindent	\V_{A,k} & \bm{V}_{AB,k} \\
		\bm{V}_{AB,k}^T & \bm{V}_{B,k}
	\end{bmatrix}, \ \bm{\bar r}_k \to \begin{bmatrix} \bm{\bar r}_{A,k} \\ \bm{\bar r}_{B,k} \end{bmatrix},
}	
where $\bm{V}_{A,k}$ is a $2(\ell-1)\times 2(\ell-1)$ matrix describing modes 1
to $\ell-1$ and $\bm{V}_{AB,k}$ is a $2(\ell-1) \times 2$ matrix describing the
correlations between modes 1 to $\ell-1$ and mode $\ell$. Finally, $\bm{V}_{B,k}$ is
a $2 \times 2$ matrix describing mode $\ell$, the mode being measured.
\item Calculate the update rules for the case where no click is detected
\sseq{
	\bm{V}_{A,k}' &\to \bm{V}_{A,k} - \bm{V}_{AB,k} (\bm{V}_{B,k}+\one_2 )^{-1} \bm{V}_{AB,k}^T,\label{Eq:MatrixUpdate}\\
	\bm{\bar r}_{A,k}'&\to\bm{\bar r}_{A,k}- \bm{V}_{AB,k} (\bm{V}_{B,k}+\one_2)^{-1} \bm{\bar r}_{B,k}.
}
\item Calculate the click probability:
\eq{ \label{prob}
	p = \sum_{k=1}^{M}  a_k q_k, \text{ with } q_k=\frac{2  e^{-\bm{\bar r}_{B,k}^T (\bm{V}_{B,k}+\one_2)^{-1} \bm{\bar r}_{B,k} }}{\sqrt{\text{det}\left( \bm{V}_{B,k} +\one_2 \right)}}.
}
\item Flip a coin with bias $p$
\item If a click is obtained, then 
\eq{
	\rho_{\ell-1} \to & \sum_{k=1}^{M} a_k \frac{\rho_{\ell-1,k}(\bm{V}_{A,k},\bm{\bar r_{A,k}}) -q_k \rho_{\ell-1,k}(\bm{V}_{A,k}',\bm{\bar r}_{A,k}') }{1-p}  
}
otherwise 
\eq{\rho_{\ell-1} \to \sum_{k=1}^{M} \left(\frac{a_k q_k}{p}  \right)  \rho_{\ell-1,k}(\bm{V}_{A,k}',\bm{\bar r_}{A,k}').
}
\item At the end of this update rule, we end up with an $\ell-1$ mode state. If
no click is recorded, the state is described as a linear combination of $M$
Gaussian states. If a click is recorded, the $\ell-1$-mode state is described
using $2M$ Gaussian states.
\end{itemize}
If the initial state is just a Gaussian state then $M=1$. As the algorithm
progresses and $m$ clicks are detected, $2^m$ Gaussian states are required to
describe the conditional state since each click doubles the number of Gaussian
states in the linear combination. 
As shown in Ref.~\cite{quesada2018gaussian}, when $m$ clicks are
detected, the complexity of the algorithm is $O(\ell^2 2^m)$ for a system of $\ell$ modes. Thus, this
algorithm is best suited to simulating settings involving a large number of
modes and comparatively few detector clicks.

\section{Benchmarking} \label{sec:bench}
\begin{figure*}[t!]
  \includegraphics[width=0.47\textwidth]{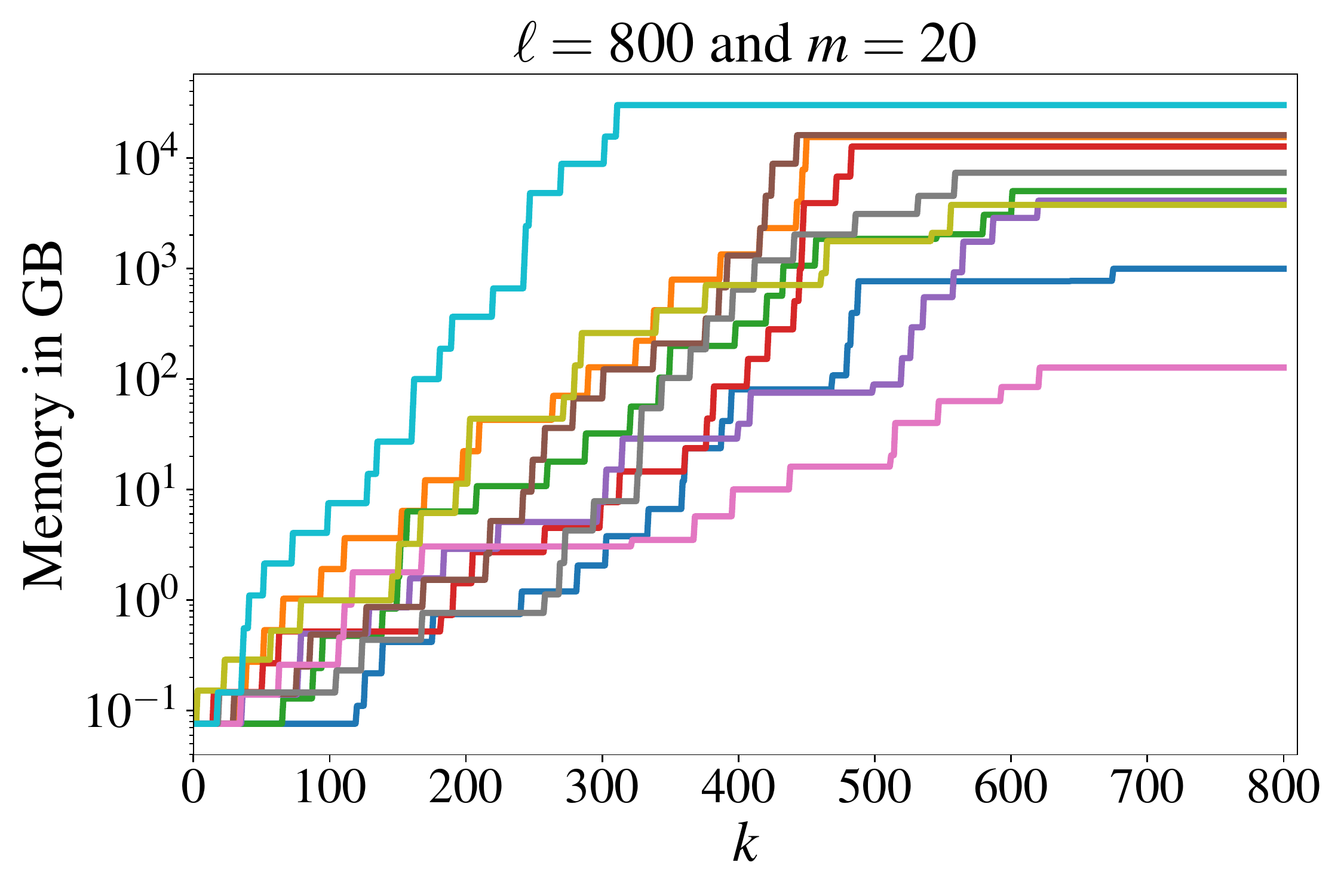}
  \includegraphics[width=0.47\textwidth]{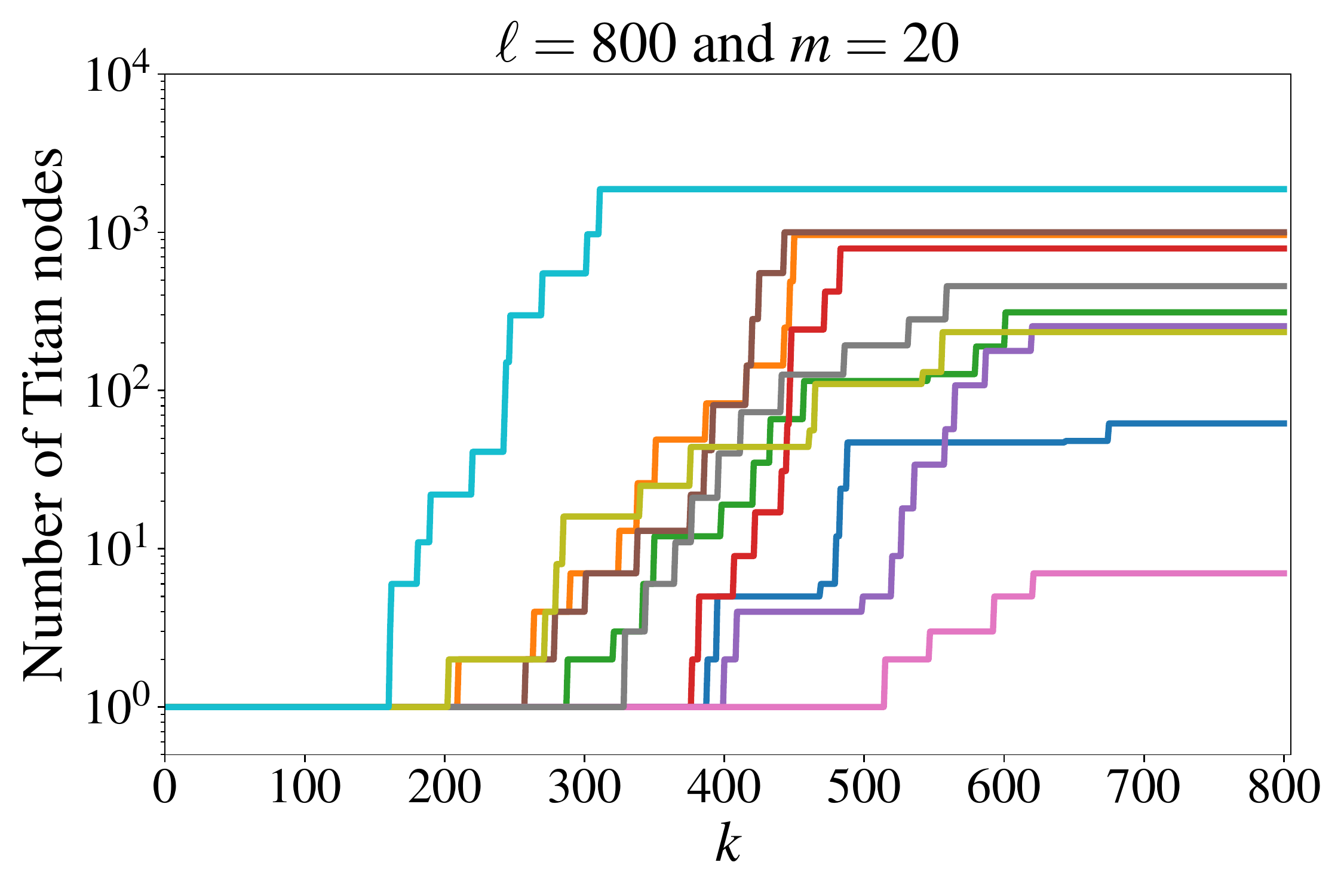}
  \caption{Memory (left panel) and Titan compute nodes (right panel) required to generate
a sample, plotted with respect to the number of steps $k$ in the algorithm for a device with $\ell=800$ modes and $m=20$ clicks. Each curve corresponds to a different
random sample and jumps in each curve mark a click. Typically $10^3-10^4$ GB of memory and $10^2-10^4$ Titan compute nodes are needed to
simulate samples with 20 clicks in a device with 800 modes.}
\label{n20ell800}
\end{figure*}
Before discussing the numerical implementation of the algorithm, we
estimate the associated computational resource requirements. Recall that the 
algorithm is based on sequential measurements, each one yielding a
new conditional state for a subsequent measurement. Therefore, advancing to the next
step requires storing the covariance matrix and the mean vector associated to
the updated conditional state as well as evaluating 
the probability of click in the next measurement. Since the number of
Gaussian states required to represent this conditional state increases
exponentially with the number of clicks observed, the memory to store them and
compute time to evaluate their sum also scales exponentially. 

All numerical computations discussed in this paper are performed using the Titan
supercomputer housed at the Oak Ridge National Laboratory. Titan is equipped
with 18,688 compute nodes, each having 16 CPU cores and 32 gigabytes of memory.
Hence with a total of 299,008 CPU cores and 598,016 gigabytes of memory it
has a 27 petaFLOPS of theoretical peak performance~\cite{titan}. In the following we discuss the memory and runtime requirements for the algorithm.  

\begin{figure*}
  \includegraphics[width=0.47\textwidth]{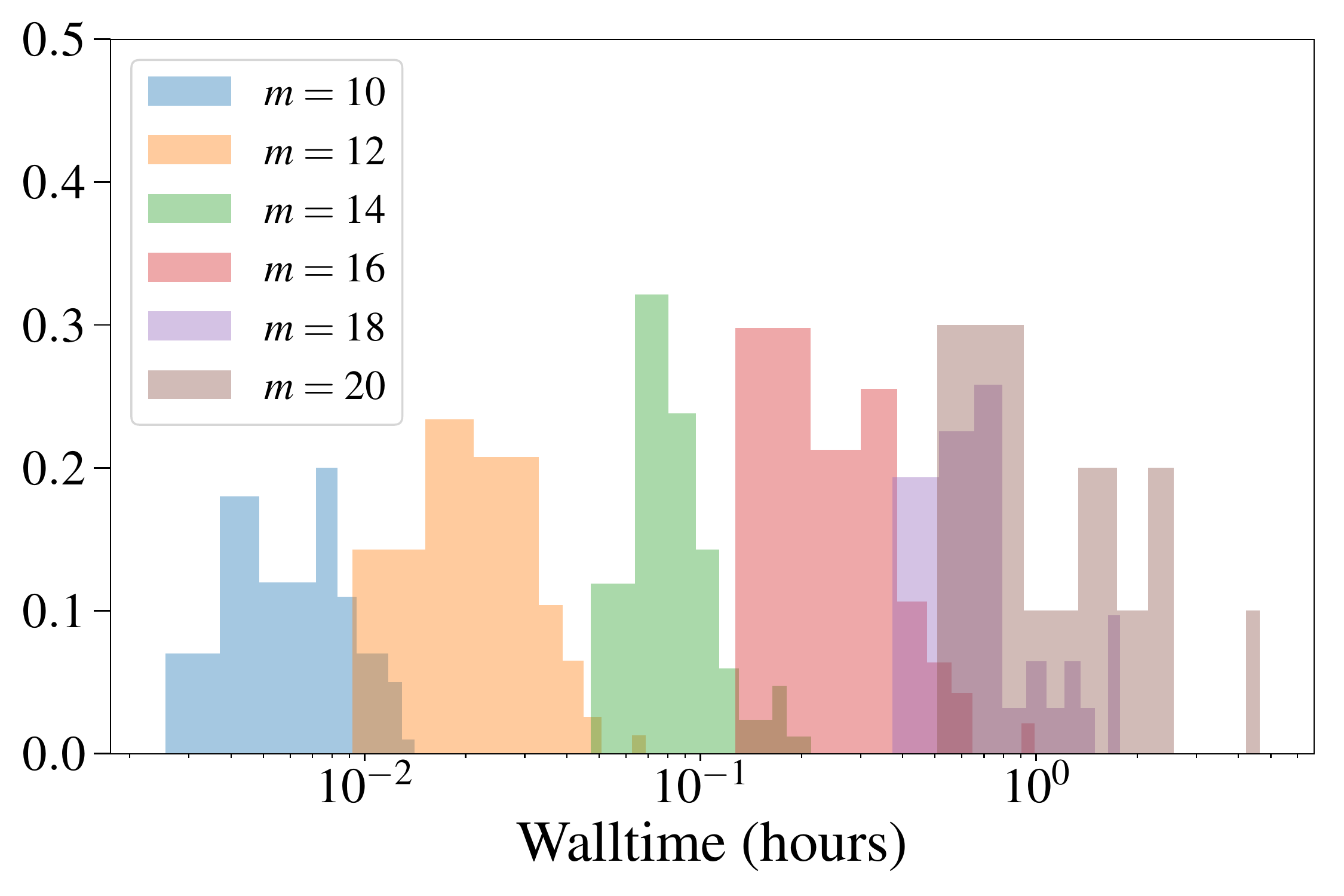}
  \includegraphics[width=0.47\textwidth]{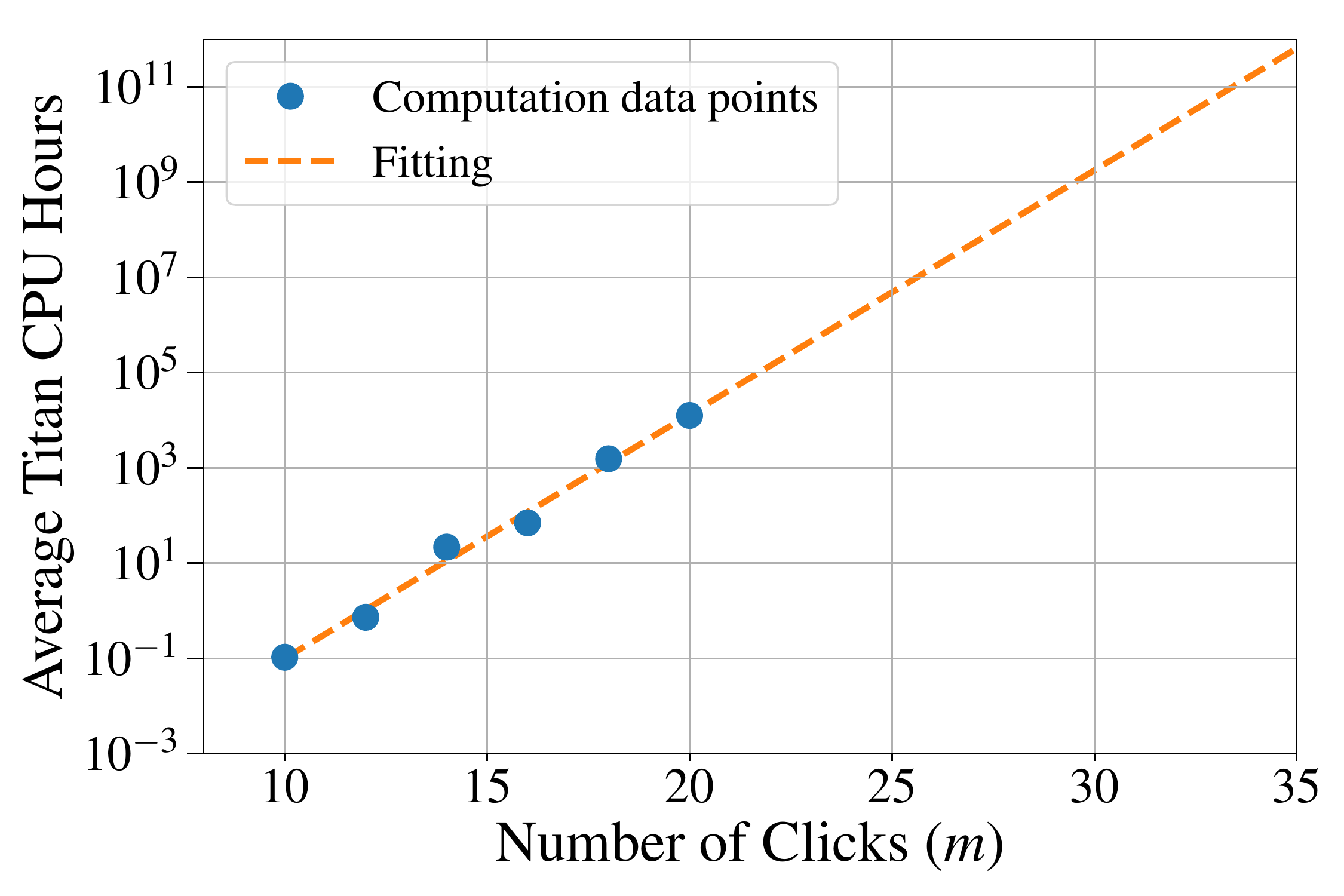}
  \caption{Histogram of walltimes (left panel) and average CPU hours (right
panel) to simulate samples with various number of clicks $m$ and modes
$\ell=2m^2$. As shown in the left panel the walltime for a given number of
clicks can vary depending on when the clicks are observed during the execution
of the algorithms. If more clicks are observed earlier in the process, the
walltime is higher. The average CPU hours, as shown in the right panel
increases exponentially. Therefore, the algorithm discussed in this paper
becomes intractable, even with supercomputers, for click sizes around $m=25$ when
$\ell=2m^2$.}
\label{avgtimewithclicks}
\end{figure*}
\subsection{Memory}

Consider an $\ell$-mode Gaussian initial state with covariance matrix
$V$  of size $2\ell\times2\ell$. For benchmarking purposes, we consider input states
with 8 dB of squeezing and a linear optical network described by unitaries drawn
from the Haar measure. For concreteness, we start the measurement at the
$\ell$-th mode. After measurement of this mode, we obtain
the conditional state of the remaining $\ell-1$ modes for the next measurement. If no click is
observed, the conditional state remains Gaussian for
the remaining modes with an updated covariance matrix of size
$2(\ell-1)\times2(\ell-1)$. On the other hand, if a click is observed, the
conditional state for the remaining modes is given by a linear combination of
two separate Gaussian states with covariance matrices of sizes
$2(\ell-1)\times2(\ell-1)$.  Hence, each time a click is observed in the
sequence of measurements, the number of covariance matrices needed to
characterize the conditional state doubles. After $m$ clicks have been
recorded at the $k$-th step, the state for the next step will be given by $2^m$
covariance matrices, each of size $2(\ell-k)\times 2(\ell-k)$. At this stage we need to store as many as $4(\ell-k)^2\times2^m$ matrix elements
to represent the state. Furthermore, numerical simulations show that to
obtain an accurate estimate of the probabilities, it is necessary to work with
\texttt{quad precision}, which requires 16 bytes of memory to store one floating
point number. Therefore, the total memory (in gigabytes) to store the covariance
matrices corresponding to the state at the $k$-th step is:
\begin{align}\label{eq:memory}
 \mathbb{M}_k(m) &= \eta\times4(\ell-k)^2\times2^m\times 16 /2^{30} ~~{\rm giga
bytes} \nonumber \\
                 &= \eta~2^{m-24} (\ell-k)^2 ~~{\rm gigabytes},
\end{align}
where $\eta\approx2$ is an overhead factor introduced to account for the
temporary variables created during the execution of the algorithm.

Eq.~\eqref{eq:memory} dictates that the memory requirement increases exponentially
with $m$. For large enough $m$ and $\ell$, a local desktop computer
is not sufficient to run the algorithm, since a large memory capacity is required. In order to overcome this issue, we use a distributed memory parallel
implementation of the algorithm where covariance matrices are allocated
over multiple compute nodes of a supercomputer
(or a compute cluster). The evaluation of the sum in Eq.~\eqref{prob} is
distributed over multiple compute nodes where each one performs a partial
sum. Subsequently, at each step of the algorithm we accumulate different partial sums 
using the message passing interface (MPI) protocol. In addition, we employ 
OpenMP multithreading for enhanced speed in evaluating the partial sum at each
node. Hence, the required computational resources in terms of number of compute
nodes is decided by the memory requirement of the problem and the available
memory at each node of the cluster. If each compute node has $\mu$
gigabytes of memory, the number $N$ of compute nodes required is at least
\begin{equation}
N = \frac{\mathbb{M}_k}{\mu},
\end{equation}
where $\mathbb{M}_k$ is the memory needed at step $k$ during the computation.
Each compute node of the Titan supercomputer has memory of $\mu = 32$ gigabytes
and 16 cores.

Let us consider a GBS device with 800 modes. Figure~\ref{n20ell800} 
shows the memory and number of Titan nodes $N$ used
as we progress through different steps of the algorithm. We consider 10 different random
samples with 20 clicks in each sample. Each time a click is observed, the number
of covariance matrices double and so does the memory needed to store them. This
is marked by jumps in each curve. A total of
roughly $10^4-10^5$ GB of memory and $10^3-10^4$ nodes, each with 16
processors, are needed to generate a typical sample with 20 clicks in an 800-mode device. 

It is noteworthy that the memory requirements are higher for samples in which 
more clicks appear in earlier steps of the algorithm. Therefore, for a
sample of $m$ clicks, the memory requirement would be highest if all clicks appear in the
first $m$ steps. Consider the following two scenarios for a system of $\ell$ modes: (i) all
$m$ clicks are observed in the last $m$ modes, and (ii) all $m$ clicks are
observed in the first $m$ modes. In the former situation, because no doubling of the number of covariance matrices occurs for the first $\ell-m$ modes, we only need to store a single covariance matrix at each step. Moreover, since the size of this matrix is equal to the number of remaining modes, it decreases with each measurement. Once clicks are detected, the number of covariance matrices increases exponentially, but they are small in size. Conversely, in the latter case, after measuring the first $m$ modes we already need to deal with an exponential number of large covariance matrices, which requires significantly more memory. 

\subsection{Runtime}

\begin{figure*}
  \includegraphics[width=0.47\textwidth]{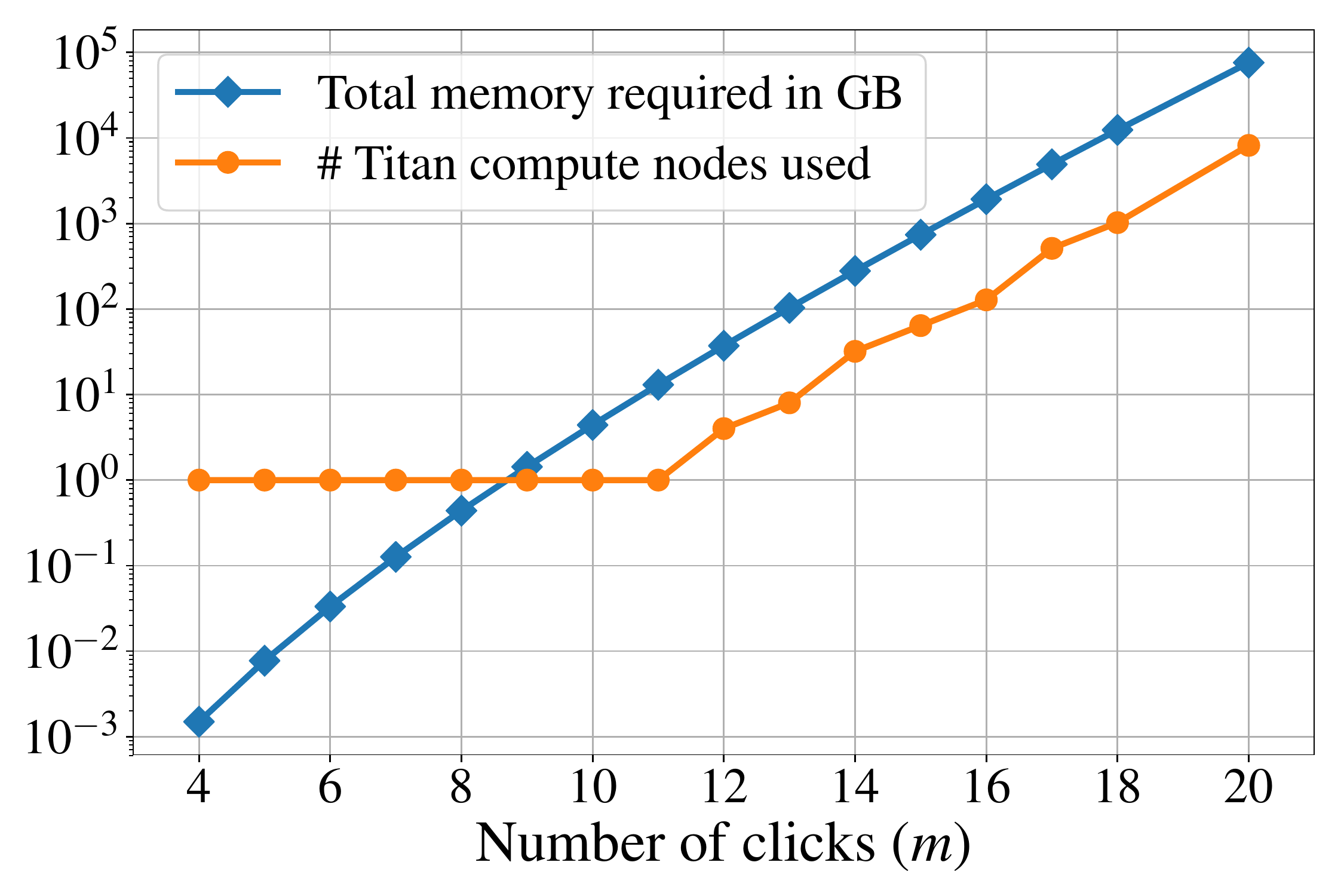}
  \includegraphics[width=0.47\textwidth]{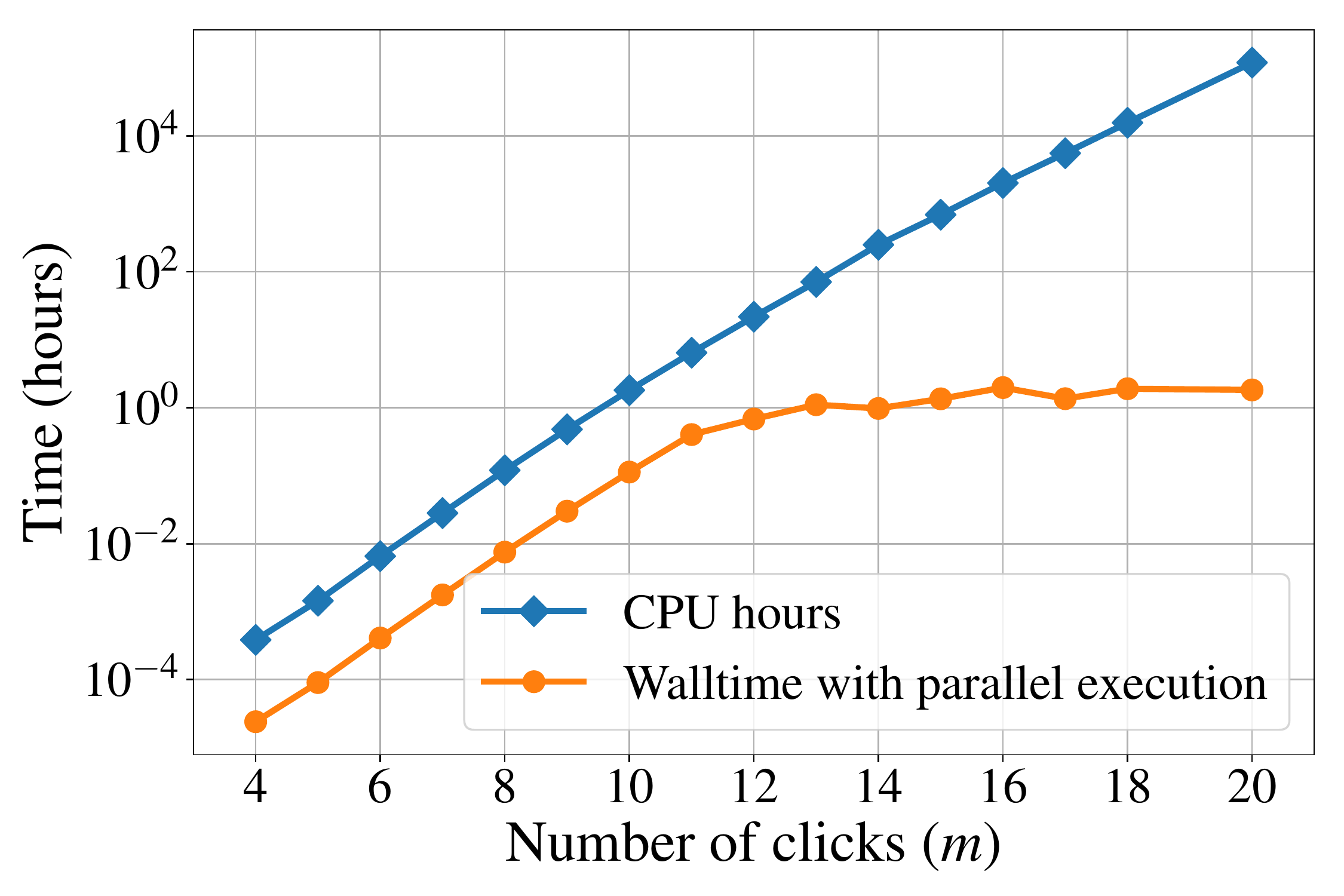}
  \caption{Requirements for memory and appropriate number of Titan nodes (left panel) as well as
associated CPU hours and walltimes for various values of $m$ (right panel), with $\ell=2m^2$,
for extreme samples where all the clicks appear in the first $m$ of $\ell$ steps
of the algorithm. It is evident from the left panel that in order to meet the
memory requirements of the simulations, an increasingly large number of Titan nodes are required
for larger $m$, i.e., they both scale exponentially with $m$. The total 
associated CPU hours (blue diamond curve in the right panel) also scales
exponentially. Using the appropriate numbers of Titan nodes, the large
simulations (with $m>12$) can be carried out in walltime of approximately two
hours. Note that there are 18,688 compute nodes available on Titan
supercomputer which can simulate these extreme samples of up to $m=22$ with
$\ell=2m^2$.}  
\label{timewithclicks}
\end{figure*}

The runtime of the algorithm is dominated by the requirement to access and process the covariance matrices stored in memory during the execution of the algorithm. Thus, runtime also scales exponentially with the number of clicks, with variations around the average runtime arising from whether most clicks are detected early or late in the algorithm. For benchmarking purposes, we fix the output in advance to have the desired number of clicks. From the perspective of the algorithm, this means that instead of flipping a coin with bias $p$ as in Eq.~\eqref{prob}, we set the outcome of this coin flip beforehand to a desired value. This allows us to more efficiently study the properties of the algorithm without the need to wait until a sample of the desired click size is randomly obtained.

Fig.~\ref{avgtimewithclicks} shows histograms of walltimes (real user time between start and end of the simulation) to obtain a sample
of various numbers of clicks $m$. Here the position of the clicks was chosen uniformly at random. 
For each value of $m$, there is a distribution
of walltimes: higher walltimes correspond to the cases for which most of the
clicks are observed earlier in the execution of the algorithm. The right panel
of Fig.~\ref{avgtimewithclicks} shows the average CPU hours as a function of number of clicks $m$. It is evident that on average
the CPU hours scale exponentially. A fit to these runtimes indicate that
simulations of samples with $m=30$ require $10^9$ CPU hours, which is
equivalent to the entire Titan supercomputer running for approximately five months,
assuming the availability of sufficient memory.

\subsection{Worst-case setting}

We now consider the memory and runtime requirements in the worst-case scenario of all $m$ clicks appearing in the first $m$ modes.
Fig.~\ref{timewithclicks} shows the computing resources needed to generate samples with different values of $m$, with $\ell=2 m^2$. Tables~\ref{tab:memory} and~\ref{tab:run} provide supporting numbers.
Note that $\ell=2 m^2$ is the regime of small collision probability~\cite{aaronson2011computational}, where the effect of threshold detectors is minor compared to photon-number-resolving detectors \cite{quesada2018gaussian}. The left panel shows the memory and Titan nodes
used for the simulations and the right panel shows the CPU hours and
the walltime when appropriate number of compute nodes are utilized for
parallel execution of the algorithm.
\begin{table}[t!]
\caption{Maximum memory needed to simulate one GBS sample for different number of modes $\ell$ and clicks $m$, where we fix $\ell=2m^2$.}
\begin{tabular}{{l@{\hskip 0.5in}c@{\hskip 0.5in}c}}
\toprule[1pt]
$\ell$ &  $m$ &  Memory (GB) \\
\hline
    50 &    5 &        0.008 \\
   200 &   10 &        4.407 \\
   450 &   15 &       739.16 \\
   800 &   20 &        76050 \\

\bottomrule[1pt]
\end{tabular}
\label{tab:memory}
\end{table}

\begin{table}[t!]
\caption{Maximum runtime to simulate one sample for a device with $\ell=2m^2$.}
\vspace{0.3cm}
\begin{tabular}{{c@{\hskip 0.2in}c@{\hskip 0.2in}c@{\hskip 0.2in}c@{\hskip 0.2in}c}}
\toprule[1pt]
$\ell$ & $m$ &       Titan nodes &           CPU hours &       Walltime (hours) \\
\hline
 200 & 10 &     1 &      1.81 &  0.11 \\
 288 & 12 &     4 &     21.86 &  0.68 \\
 392 & 14 &    32 &    250.17 &  0.97 \\
 512 & 16 &   128 &   2,028.91 &  1.98 \\
 624 & 18 &  1024 &  15,612.79 &  1.90 \\
 800 & 20 &  8192 &  239,773.95 &  1.83 \\
\bottomrule[1pt]
\label{tab:run}
\end{tabular}
\end{table}
It is evident from the left panel that in
order to meet the memory requirements of the simulations, an exponentially large number of compute nodes are needed. The walltime for running the
algorithm can be brought down to approximately two hours by using the appropriate
number of compute nodes. 
For $m=20$ and $\ell=800$, a total of 8192 Titan nodes (131,072 processors) were employed -- which is approximately 40\% of the available CPUs on Titan -- running for two hours. Hence, a total of 240,000 CPU hours and approximately 100,000 gigabytes of
memory were consumed (see Table \ref{tab:run}). Therefore, a simulation of this algorithm with $m>22$ and $\ell=2m^2$ would be beyond the
current capabilities the Titan supercomputer. 

In the following section, we use the algorithm to generate samples for manageable
problem sizes and study the application of these samples to dense
subgraph identification.

\section{Application: Dense subgraph identification} \label{sec:app}

We study an NP-Hard optimization task known as the densest $k$-subgraph problem:
given a graph $G$ with $n$ vertices, find the subgraph of size $k<n$ with the
largest number of edges \cite{feige2001dense}. This problem has a connection to
clustering tasks that aim at finding highly correlated subsets of data, with
applications in a wide range of fields such as as data mining
\cite{kumar1999trawling,angel2012dense,beutel2013copycatch,chen2012dense},
bioinformatics \cite{fratkin2006motifcut,saha2010dense}, and finance
\cite{arora2011computational}.

It was shown in Refs.~\cite{arrazola2018using, arrazola2018quantum} that GBS can
be employed to enhance classical algorithms for the densest $k$-subgraph
problem. The main insight of this approach is that, using the encoding technique
of Ref.~\cite{bradler2017gaussian}, a GBS device can be configured to sample
subgraphs in such a way that the probability of observing each subgraph is directly correlated with its density. This leads to a sampling device that
selects dense subgraphs with high probability, thus aiding in their
identification. In this setting, the number of modes is equal to the number of
vertices in the graph and subgraphs are identified by postselecting on
collision-free samples and assigning vertices corresponding to the modes where clicks were detected.

Classical simulation algorithms can be used to quantitatively benchmark the
improvement that can be attained when using GBS to find dense subgraphs.
Previous simulation techniques were limited to graphs of a few dozen vertices
and were incapable of including the effects of experimental imperfections such
as photon loss \cite{arrazola2018using}. Here, we employ the algorithm of Ref.~\cite{quesada2018gaussian}
to simulate threshold GBS for a variety of instances.

\begin{figure}[t!]
  \includegraphics[width=0.47\textwidth]{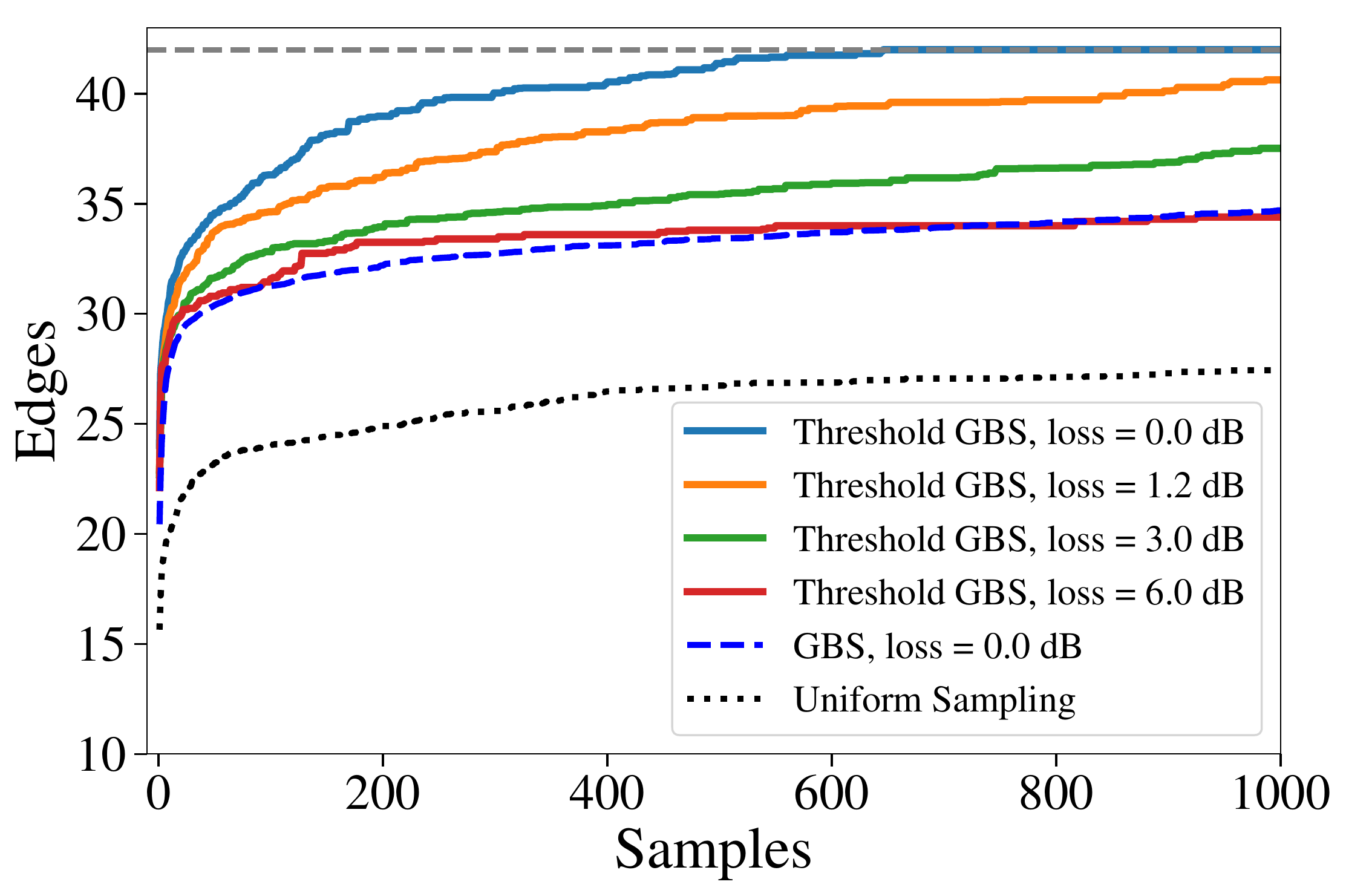}
  \caption{Using random search to find a dense $10$-vertex subgraph of the planted $30$ vertex graph
in Sec.~\ref{Sec:Planted}. Here, the number of edges in the
densest found subgraph is plotted against the number of samples. Each line
corresponds to taking samples from a different underlying distribution, as
summarized by the legend. Results are averaged over a minimum of $20$ runs to
reduce statistical effects. We see that, in the loss-free regime, random search
using threshold GBS outperforms the strategy of postselecting on collision-free
outputs from PNRs in conventional GBS, being able to find the densest subgraph
($42$ edges - dashed black horizontal line) after approximately $600$ samples.
Moreover, threshold GBS is still advantageous in the lossy regime, outperforming
uniform random sampling for all loss levels considered and even performing as well as loss-free GBS.}\label{applicationfig3010}
\end{figure}

\subsection{Random graph with planted subgraph}\label{Sec:Planted}
We study the random graph of 30 vertices presented in Ref.~\cite{arrazola2018using}, which is built as follows. First, a
random graph of 20 vertices was created, where each edge was added with probability $p=0.5$. Second, a planted random graph of 10 vertices was constructed with
probability $q=0.875$ of adding each edge. Finally, eight vertices were selected
uniformly at random in both graphs and an edge was added between them to define a full
graph of 30 vertices. This resulting graph has the property that the planted
subgraph has the largest density among subgraphs of size 10, even though its
vertices have a smaller degree than the average for the full graph. This makes
the densest subgraph difficult to identify for algorithms based on vertex
degree. 

We study the performance of random search algorithms for identifying the planted
subgraph, where the strategy is to randomly sample subgraphs and select the
densest among all outputs. We compare three different strategies: (i) uniform
sampling, (ii) GBS postselected on collision-free outcomes, i.e., outcomes where
not more than one photon is detected in each mode, and (iii) GBS with threshold
detectors. For threshold GBS, we also investigate the effect of losses, sampling with loss values of $1.2$dB ($25 \%$), $3.0$dB ($50
\%$), and $6.0$dB ($75 \%$). Note that the ability to simulate losses is a unique feature of the algorithm of Ref.~\cite{quesada2018gaussian}.

The results are summarized in Fig.~\ref{applicationfig3010}. As expected, both
loss-free versions of GBS with photon-number-resolving (PNR) and threshold detectors
outperform uniform sampling. Interestingly, GBS with threshold detectors
performs noticeably better than PNR-based GBS with collision-free
postselection. Indeed, threshold GBS consistently finds the planted graph after a few hundred
samples. The effect of a threshold detector is equivalent to a post-processing
of the PNR outputs where events with two or more photons detected in a mode are
considered as a single click in those modes. Thus, our results indicate that
retaining collision outputs is beneficial for the purpose of dense subgraph
identification, a task that is done automatically by threshold detectors.
Moreover, a performance advantage persists for threshold GBS even in the lossy regime. Here, even with a loss of $6.0$dB, threshold GBS is comparable to PNR-based GBS with postselection
and no losses.

In this example, we are interested in finding the densest subgraph of 10 vertices. The proportion of $10$ click samples depends on the
amount of loss, and we postselect on $4\times10^6$ samples yielding: $4.1 \%$ of
samples for no losses, $3.2 \%$ for $1.2$ dB loss, $2.0\%$ for $3.0$ dB loss,
and $0.5\%$ for $6.0$ dB loss. Simulation of each of these cases needed approximately 1000 CPU
hours and was completed in less than half an hour using 150 Titan compute nodes
each having 16 CPU cores.

\begin{figure}[t!]
  \includegraphics[width=0.47\textwidth]{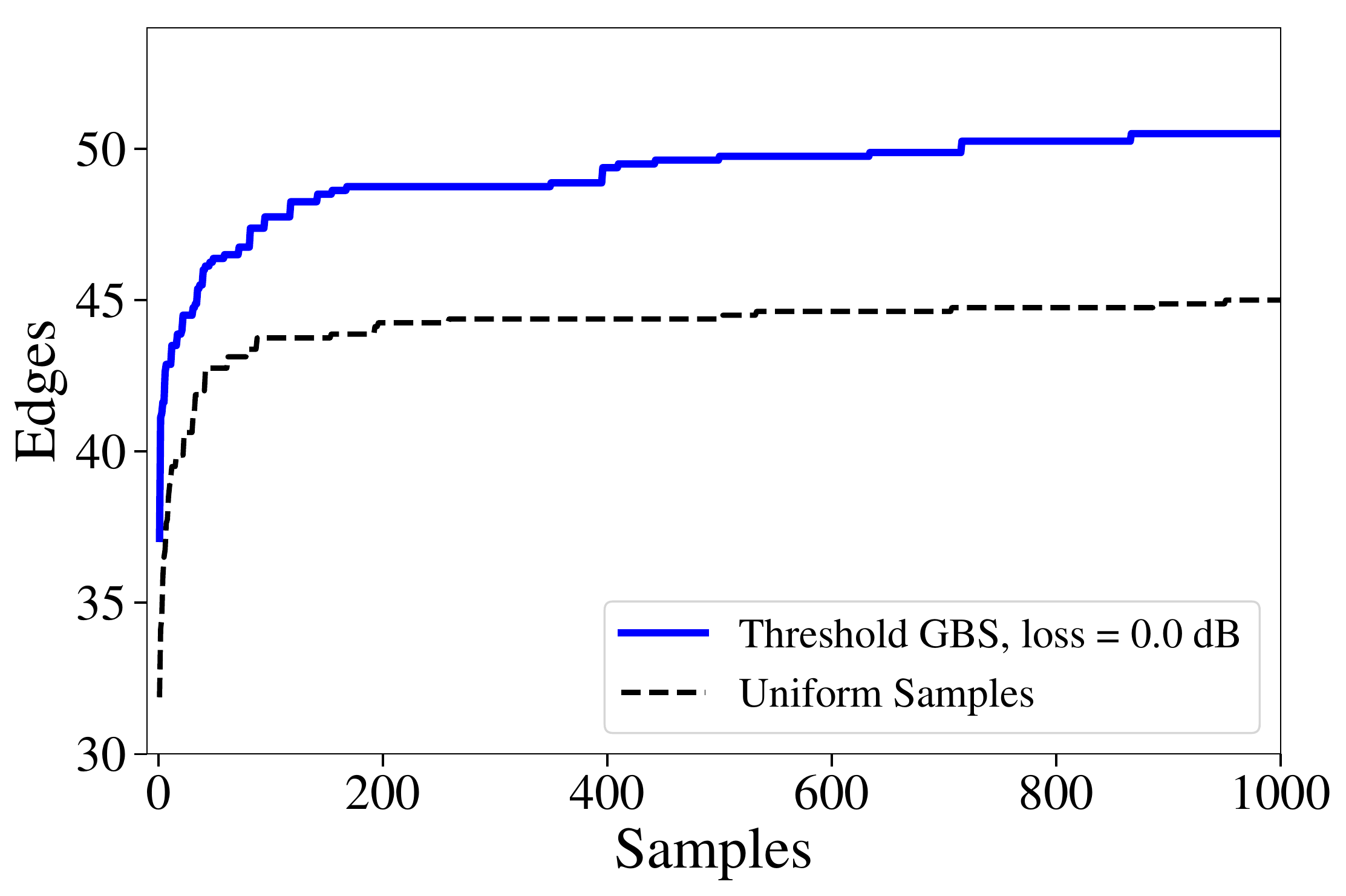}
  \caption{Enhancing random search using threshold GBS for dense subgraph
identification in the $\textsf{brock200\textunderscore 2}$
graph~\cite{johnson1996cliques} of Sec.~\ref{Sec:DIMACS}. Here, the number of
edges in the densest found $12$ vertex subgraph is plotted against the number of
samples, and lines correspond to sampling from different distributions. Results
are averaged over $10$ independent runs. We see that samples from threshold GBS
(solid blue line) allow random search to outperform uniform sampling (dashed
gray line). The densest subgraph of size $12$ is known to be complete, with $66$
edges, motivating the use of more advanced heuristics and the combination with
threshold GBS. It took $30$ gigabytes of memory for each samples, and in total
45,000 core hours to generate 260,000 samples of which we postselect
approximately 10,000 samples with 12 clicks.}\label{applicationfig20012}
\end{figure}

\subsection{DIMACS graph}
\label{Sec:DIMACS}
 
We now investigate the performance of GBS with threshold detectors for dense
subgraph identification for a much larger graph of $200$ vertices. We use the
$\textsf{brock200\textunderscore 2}$ graph~\cite{johnson1996cliques} of the DIMACS dataset and search
for the densest subgraph with $12$ vertices. The maximum clique, i.e., the
largest complete subgraph, of this graph is known to be dimension $12$, meaning
that the densest subgraph has $66$ edges. For a graph of
this size, it is intractable to perform a brute force GBS postselected on
collision-free outcomes. 

We use approximately $30$ gigabytes of memory for each sample and a total of
$45,000$ core hours to generate $260,000$ samples. Out of these, $9000$ samples with $12$ clicks were postselected , which we use to compare the
performance of random search using GBS with threshold detectors to uniform
random search. The results are shown in Fig.~\ref{applicationfig20012}. Here it
can also be seen that, as expected, threshold GBS enhances random search. This
simple sampling strategy, however, is unable to find the optimum subgraph of
$66$ vertices, showcasing the increased difficulty of dense subgraph
identification for large graphs. In such cases, it is preferable to combine GBS
with more advanced heuristics~\cite{arrazola2018using}.

\section{Conclusion} \label{sec:disc}
Our results constitute the largest simulation of GBS with threshold detectors to
date. We have thus set a first goalpost for experiments aiming to build quantum devices
capable of outperforming classical algorithms: they must be able to generate
enough squeezing to observe more than 20 clicks on average over systems of
several hundred modes. The difficulty of classically simulating GBS depends
both on the expected number of clicks and the number of modes. The number of
clicks in our simulations could thus have in principle been increased at the expense
of a smaller number of modes. Regardless, fixing the number of modes as $\ell=2m^2$, simulating a sample with $m>22$ clicks is beyond the current capabilities of the Titan supercomputer.

The simulation algorithm of Ref.~\cite{quesada2018gaussian} is ideally suited for
benchmarking near-term devices. First, it is an exact sampling algorithm: it
generates samples from the GBS distribution, not an approximate one. Second,
the algorithm can natively incorporate the effect of experimental imperfections,
making it ideal to verify that the physical devices are working adequately and to
test the performance of GBS for practical applications,
namely identifying dense subgraph.

Currently, the bottleneck of the classical algorithm presented here is the
memory usage, which restricts the maximum number of clicks that can be simulated
for a given number of modes. Finding more efficient methods of memory
allocation or altogether new algorithms is necessary to push simulation
capabilities even further. Indeed, it is an important open question whether
existing algorithms for boson sampling with indistinguishable single photons can
be extended to GBS, where squeezed vacuum inputs are used. These techniques
include an approximate algorithm \cite{neville2017classical} and an exact sampling algorithm \cite{clifford2018classical}.
The complexity of these algorithms is determined by hardness of computing the
permanent of a matrix of size given by the number of photons and using
supercomputers one can simulate a boson sampling event for approximately 50
photons \cite{neville2017classical,wu2016computing}. For GBS, the complexity is
determined by the hafnian of matrices with sizes equal to the number of
clicks in the sample. Based on the results reported in
Ref.~\cite{bjorklund2018faster}, computing the hafnian of a $54\times 54$ matrix
requires approximately 1000 seconds on a supercomputer. Therefore, if the
techniques of Refs. \cite{neville2017classical,clifford2018classical} could be
extended to GBS, limits of simulations presented here 
can be potentially pushed further. So far such an extension is still an
open problem.

A python version of the code designed to run on local desktop computers is
available at \cite{torcode}. Optimized code to run large simulations on a
supercomputer can be made available upon written request to the authors. 

\section*{Acknowledgements}
The authors thank Nathan Killoran, Joshua Izaac, Patrick Rebentrost,
and Christian Weedbrook for useful discussions and valuable feedback. This
research used resources of the Oak Ridge Leadership Computing Facility at the
Oak Ridge National Laboratory, which is supported by the Office of Science of
the U.S.  Department of Energy under Contract No. DE-AC05-00OR22725.


\begingroup\begin{thebibliography}{39}
\expandafter\ifx\csname natexlab\endcsname\relax\def\natexlab#1{#1}\fi

\bibitem{preskill2018quantum}
J.~Preskill, ``Quantum {C}omputing in the {NISQ} era and beyond'', {\em
  {Quantum}} {\bfseries 2} (2018) 79.

\bibitem{harrow2017quantum}
A.~W. Harrow and A.~Montanaro, ``Quantum computational supremacy'', {\em
  Nature} {\bfseries 549} (2017), no.~7671, 203.

\bibitem{pednault2017breaking}
E.~Pednault, J.~A. Gunnels, G.~Nannicini, L.~Horesh, T.~Magerlein,
  E.~Solomonik, and R.~Wisnieff, ``Breaking the 49-qubit barrier in the
  simulation of quantum circuits'', {\em arXiv:1710.05867}, 2017.

\bibitem{chen201864}
Z.-Y. Chen, Q.~Zhou, C.~Xue, X.~Yang, G.-C. Guo, and G.-P. Guo, ``64-qubit
  quantum circuit simulation'', {\em Science Bulletin}, 2018.

\bibitem{zulehner2018advanced}
A.~Zulehner and R.~Wille, ``Advanced simulation of quantum computations'', {\em
  IEEE Transactions on Computer-Aided Design of Integrated Circuits and
  Systems}, 2018.

\bibitem{biamonte2018quantum}
J.~D. Biamonte, M.~E. Morales, and D.~E. Koh, ``Quantum supremacy lower bounds
  by entanglement scaling'', {\em arXiv:1808.00460}, 2018.

\bibitem{chen2018classical}
J.~Chen, F.~Zhang, M.~Chen, C.~Huang, M.~Newman, and Y.~Shi, ``Classical
  simulation of intermediate-size quantum circuits'', {\em arXiv:1805.01450},
  2018.

\bibitem{aaronson2011computational}
S.~Aaronson and A.~Arkhipov, ``The computational complexity of linear optics'',
  in ``Proceedings of the forty-third annual ACM symposium on Theory of
  computing'', pp.~333--342, ACM.
\newblock 2011.

\bibitem{spring2012boson}
J.~B. Spring, B.~J. Metcalf, P.~C. Humphreys, W.~S. Kolthammer, X.-M. Jin,
  M.~Barbieri, A.~Datta, N.~Thomas-Peter, N.~K. Langford, D.~Kundys, {\em
  et~al.}, ``Boson sampling on a photonic chip'', {\em Science}, 2012 1231692.

\bibitem{broome2013photonic}
M.~A. Broome, A.~Fedrizzi, S.~Rahimi-Keshari, J.~Dove, S.~Aaronson, T.~C.
  Ralph, and A.~G. White, ``Photonic boson sampling in a tunable circuit'',
  {\em Science} {\bfseries 339} (2013), no.~6121, 794--798.

\bibitem{tillmann2013experimental}
M.~Tillmann, B.~Daki{\'c}, R.~Heilmann, S.~Nolte, A.~Szameit, and P.~Walther,
  ``Experimental boson sampling'', {\em Nature Photonics} {\bfseries 7} (2013),
  no.~7, 540.

\bibitem{aaronson2013bosonsampling}
S.~Aaronson and A.~Arkhipov, ``Bosonsampling is far from uniform'', {\em
  arXiv:1309.7460}, 2013.

\bibitem{lund2014boson}
A.~Lund, A.~Laing, S.~Rahimi-Keshari, T.~Rudolph, J.~L. O’Brien, and
  T.~Ralph, ``Boson sampling from a gaussian state'', {\em Physical Review
  Letters} {\bfseries 113} (2014), no.~10, 100502.

\bibitem{bentivegna2015experimental}
M.~Bentivegna, N.~Spagnolo, C.~Vitelli, F.~Flamini, N.~Viggianiello,
  L.~Latmiral, P.~Mataloni, D.~J. Brod, E.~F. Galv{\~a}o, A.~Crespi, {\em
  et~al.}, ``Experimental scattershot boson sampling'', {\em Science
  {A}dvances} {\bfseries 1} (2015), no.~3, e1400255.

\bibitem{latmiral2016towards}
L.~Latmiral, N.~Spagnolo, and F.~Sciarrino, ``Towards quantum supremacy with
  lossy scattershot boson sampling'', {\em New Journal of Physics} {\bfseries
  18} (2016), no.~11, 113008.

\bibitem{hamilton2017gaussian}
C.~S. Hamilton, R.~Kruse, L.~Sansoni, S.~Barkhofen, C.~Silberhorn, and I.~Jex,
  ``Gaussian boson sampling'', {\em Physical Review Letters} {\bfseries 119}
  (2017), no.~17, 170501.

\bibitem{kruse2018detailed}
R.~Kruse, C.~S. Hamilton, L.~Sansoni, S.~Barkhofen, C.~Silberhorn, and I.~Jex,
  ``A detailed study of gaussian boson sampling'', {\em arXiv:1801.07488},
  2018.

\bibitem{huh2015boson}
J.~Huh, G.~G. Guerreschi, B.~Peropadre, J.~R. McClean, and A.~Aspuru-Guzik,
  ``Boson sampling for molecular vibronic spectra'', {\em Nature Photonics}
  {\bfseries 9} (2015), no.~9, 615.

\bibitem{clements2017experimental}
W.~R. Clements, J.~J. Renema, A.~Eckstein, A.~A. Valido, A.~Lita, T.~Gerrits,
  S.~W. Nam, W.~S. Kolthammer, J.~Huh, and I.~A. Walmsley, ``Experimental
  quantum optical approximation of vibronic spectroscopy'', {\em
  arXiv:1710.08655}, 2017.

\bibitem{sparrow2018simulating}
C.~Sparrow, E.~Mart{\'\i}n-L{\'o}pez, N.~Maraviglia, A.~Neville, C.~Harrold,
  J.~Carolan, Y.~N. Joglekar, T.~Hashimoto, N.~Matsuda, J.~L. O’Brien, {\em
  et~al.}, ``Simulating the vibrational quantum dynamics of molecules using
  photonics'', {\em Nature} {\bfseries 557} (2018), no.~7707, 660.

\bibitem{arrazola2018using}
J.~M. Arrazola and T.~R. Bromley, ``Using gaussian boson sampling to find dense
  subgraphs'', {\em Phys. Rev. Lett.} {\bfseries 121} Jul (2018) 030503.

\bibitem{arrazola2018quantum}
J.~M. Arrazola, T.~R. Bromley, and P.~Rebentrost, ``Quantum approximate
  optimization with gaussian boson sampling'', {\em Phys. Rev. A} {\bfseries
  98} Jul (2018) 012322.

\bibitem{bradler2017gaussian}
K.~Br{\'a}dler, P.-L. Dallaire-Demers, P.~Rebentrost, D.~Su, and C.~Weedbrook,
  ``Gaussian boson sampling for perfect matchings of arbitrary graphs'', {\em
  arXiv:1712.06729}, 2017.

\bibitem{neville2017classical}
A.~Neville, C.~Sparrow, R.~Clifford, E.~Johnston, P.~M. Birchall, A.~Montanaro,
  and A.~Laing, ``Classical boson sampling algorithms with superior performance
  to near-term experiments'', {\em Nature Physics} {\bfseries 13} (2017),
  no.~12, 1153.

\bibitem{clifford2018classical}
P.~Clifford and R.~Clifford, ``The classical complexity of boson sampling'', in
  ``Proceedings of the Twenty-Ninth Annual ACM-SIAM Symposium on Discrete
  Algorithms'', pp.~146--155, SIAM.
\newblock 2018.

\bibitem{quesada2018gaussian}
N.~Quesada, J.~M. Arrazola, and N.~Killoran, ``Gaussian boson sampling using
  threshold detectors'', {\em arXiv:1807.01639}, 2018.

\bibitem{johnson1996cliques}
D.~S. Johnson and M.~A. Trick, ``Cliques, coloring, and satisfiability: second
  dimacs implementation challenge, october 11-13, 1993'', American Mathematical
  Society, 1996.

\bibitem{titan}
{Oak Ridge National Laboratory}.
  \url{https://www.olcf.ornl.gov/olcf-resources/compute-systems/titan/}.

\bibitem{feige2001dense}
U.~Feige, D.~Peleg, and G.~Kortsarz, ``The dense k-subgraph problem'', {\em
  Algorithmica} {\bfseries 29} (2001), no.~3, 410--421.

\bibitem{kumar1999trawling}
R.~Kumar, P.~Raghavan, S.~Rajagopalan, and A.~Tomkins, ``Trawling the web for
  emerging cyber-communities'', {\em Computer Networks} {\bfseries 31} (1999),
  no.~11-16, 1481--1493.

\bibitem{angel2012dense}
A.~Angel, N.~Sarkas, N.~Koudas, and D.~Srivastava, ``Dense subgraph maintenance
  under streaming edge weight updates for real-time story identification'',
  {\em Proceedings of the VLDB Endowment} {\bfseries 5} (2012), no.~6,
  574--585.

\bibitem{beutel2013copycatch}
A.~Beutel, W.~Xu, V.~Guruswami, C.~Palow, and C.~Faloutsos, ``Copycatch:
  stopping group attacks by spotting lockstep behavior in social networks'', in
  ``Proceedings of the 22nd international conference on World Wide Web'',
  pp.~119--130, ACM.
\newblock New York, 2013.

\bibitem{chen2012dense}
J.~Chen and Y.~Saad, ``Dense subgraph extraction with application to community
  detection'', {\em IEEE Transactions on Knowledge and Data Engineering}
  {\bfseries 24} (2012), no.~7, 1216--1230.

\bibitem{fratkin2006motifcut}
E.~Fratkin, B.~T. Naughton, D.~L. Brutlag, and S.~Batzoglou, ``Motifcut:
  regulatory motifs finding with maximum density subgraphs'', {\em
  Bioinformatics} {\bfseries 22} (2006), no.~14, e150--e157.

\bibitem{saha2010dense}
B.~Saha, A.~Hoch, S.~Khuller, L.~Raschid, and X.-N. Zhang, ``Dense subgraphs
  with restrictions and applications to gene annotation graphs'', in ``Annual
  International Conference on Research in Computational Molecular Biology'',
  pp.~456--472, Springer.
\newblock Berlin, 2010.

\bibitem{arora2011computational}
S.~Arora, B.~Barak, M.~Brunnermeier, and R.~Ge, ``Computational complexity and
  information asymmetry in financial products'', {\em Communications of the
  ACM} {\bfseries 54} (2011), no.~5, 101--107.

\bibitem{wu2016computing}
J.~Wu, Y.~Liu, B.~Zhang, X.~Jin, Y.~Wang, H.~Wang, and X.~Yang, ``Computing
  permanents for boson sampling on tianhe-2 supercomputer'', {\em arXiv
  preprint arXiv:1606.05836}, 2016.

\bibitem{bjorklund2018faster}
A.~Bj{\"o}rklund, B.~Gupt, and N.~Quesada, ``A faster hafnian formula for
  complex matrices and its benchmarking on the titan supercomputer'', {\em
  arXiv:1805.12498}, 2018.

\bibitem{torcode}
B.~Gupt, ``Torontonian sampling code''.
  \url{https://github.com/XanaduAI/torontonian-sampling}, 2018.

\end{thebibliography}\endgroup

\end{document}